\documentclass[12pt]{article}

\begin{document}

\author{A.I.Volokitin$^{1,2}$\footnote{Corresponding author.
\textit{E-mail address}:avoli@samgtu.ru} \hskip2mm   and B.N.J.Persson$^1$ \\
$^1$Institut f\"ur Festk\"orperforschung, Forschungszentrum \\
J\"ulich, D-52425, Germany\\
$^2$Samara State Technical University, 443100 Samara,\\
Russia}
\title{Van der Waals Frictional Drag induced by Liquid Flow in Low-
Dimensional Systems }
\maketitle

\begin{abstract}
We study the van der Waals frictional drag force induced by liquid flow 
in low-dimensional systems (2D and 1D electron systems, and 2D and 1D 
channels with liquid). We find that for both 1D and 2D systems, the frictional drag force induced by 
liquid flow may be  several orders of magnitude larger than the frictional drag 
induced by electronic current.

\end{abstract}

A great deal of attention has been devoted to the problem of frictional drag 
in low-dimensional systems 
\cite{Kumar1,Kumar2,Das,Kral,Persson7,Gramila1,Sivan} 
because of it importance for nanoscale detectors. Such  detectors would be of 
great interest in micromechanical and biological applications  
\cite{Schasfoort,Munro}, where local 
dynamical effects are intensively studied.    

A frictional drag  will act on a  2D-electron system if  an electric current flows in a second 
parallel 2D electron- system. This drag effect, suggested many years ago \cite{Pogrebinskii,Price}, has been studied  
in two-dimensional quantum wells \cite{Gramila1,Sivan}. Experiments \cite{Gramila1} show that, at least for small separations, the friction drag 
 can be explained by the interaction between the electrons in the different layers via the fluctuating Coulomb field. However for large 
inter-layer separation 
the friction drag is dominated by phonon exchange \cite{Bonsager}.      

Recently,  flow  of liquids over bundles of single-walled carbon nanotubes (SWNT)  
 was found to generate a voltage in the sample in the flow direction \cite{Kumar1,Kumar2}. The dependence of the voltage on the flow speed 
was found to be  logarithmic  over five decades of variation of the speed. There have been attempts to 
explain  this flow-induced voltage in electrokinetic terms, as a result of the streaming potential that develops along the flow of an electrolyte through 
a microporous insulator \cite{Cohen,Ghosh}. Earlier Kr\'al and Shapiro \cite{Kral} proposed that the  liquid flow transfer momentum to 
the acoustic 
phonons of the nanotube, and that the resulting  ``phonon wind'' drives  an  electric  current  in the nanotube. They also suggested, qualitatively, 
that the fluctuating Coulomb field of the ions in the liquid could drag directly the carriers in the nanotube. However, the first mechanism \cite{Kral} 
requires an enormous pressure  \cite{Kumar2}, while the second mechanism \cite{Kral} result in a  very small current, of order femtoAmperes \cite{Kumar2}. 
In 
\cite{Kumar2} another  mechanism was proposed, which is related to the second idea of Ref. \cite{Kral}, but which 
requires neither localization of carriers 
nor drag at 
the same speed as the ions: Thermal fluctuations in the ionic charge density in the fluid near the nanotube produce a stochastic Coulomb field 
 acting on the 
carries in the nanotube. The fluctuation-dissipation theorem  tells us that the zero-frequency friction coefficient is proportional to the 
time integral of the force-force correlation function,
 where the total force is 
determined  by  summation of all the forces on all charged carries in the nanotube. In \cite{Kumar2} it was assumed that the forces on the different 
charged carriers are uncorrelate, which is justified only for very low  carrier density. In fact,  \cite{Kumar2}  considered
the friction  between  a moving point charge and the surrounded medium. An external  charge will induce an ``image'' charge in the surrounded medium. 
Because of
the finite response time this ``image'' charge  lag behind the moving charge, which results in a force acting on the charge, referred to
as  the \textit{electrostatic} friction.
For neutral systems, such as a nanotube, the 
\textit{electrostatic} friction proposed in \cite{Kumar2} will  vanishing. 

In \cite{Persson7} it was assumed that the liquid molecules nearest to the nanotube form a 2D-solidlike monolayer, 
pinned to the nanotube 
by  adsorbed ions. As the liquid flow, the adsorbed solid monolayer performs stick-slip type of sliding motion along the nanotube. The drifting adsorbed ions will 
produce a voltage in the nanotube through electronic friction against free electrons inside the nanotube. As in \cite{Kral} it was assumed that the drift 
velocity of the electrons in the nanotubes is equal to the liquid flow velocity.  

In \cite{Das} a model calculations of the frictional drag were presented involving  a channel containing  overdamped Brownian particles. 
The channel  was imbedded in a wide chamber containing the same type of Brownian particles with drift velocity 
 parallel to the channel. It was found that the flow of particles in the chamber induces 
a drift of the particles in the channel.

In this Letter we will  address the intriguing idea of using frictional drag as a noncontact means to detect motion in surrounded liquid. 
We present a frictional drag theory based on the 
theory of the van der Waals friction \cite{Volokitin6,Volokitin10,RMP07}. 
The origin of the van der Waals friction is connected with the fluctuating electromagnetic field  which is always 
present outside of any medium due to thermal and quantum fluctuations of the charge 
density inside the medium.  This fluctuating electromagnetic field induces  polarization of the medium, and is responsible for many important 
phenomena such as  radiative heat transfer and the van der Waals interaction 
\cite{RMP07}.
  When two media are in relative motion, the induced polarization will lag 
behind the fluctuating polarization inducing it, and this gives rise  to the so-called  van der Waals friction. 

The origin of 
the van der Waals friction is  closely connected with the Doppler effect.  Let us consider two flat parallel surfaces, 
separated by a sufficiently
wide insulator gap, which prevents particles from tunneling across
it. If the charge carriers inside the volumes restricted by these 
surfaces are in relative motion (velocity $v$) a
frictional stress will act between surfaces. This frictional stress is
related with an asymmetry of the reflection amplitude along the
direction of motion; see Fig. 1. If one body emits radiation, then
in the rest reference frame of the second body these waves are
Doppler shifted which will result in different reflection
amplitudes. The same is true for radiation emitted by the second
body. The exchange of ``Doppler-shifted-photons'' will result in momentum 
transfer  and to the van der Waals friction. According to \cite{Volokitin6,Volokitin10}, the frictional stress between two flat parallel surfaces at separation $d\ll 
\lambda_T=c\hbar/k_BT$ is determined by 

\[
\sigma_{\|} =\frac \hbar {2\pi ^3}\int_{-\infty }^\infty dq_y\int_0^\infty
dq_xq_xe^{-2qd}\left\{ \int_0^\infty d\omega [n(\omega )-n(\omega
+q_xv)]\right.
\]
\[
\times \left( \frac{\mathrm{Im}R_{1p}(\omega +q_xv)\mathrm{Im}R_{2p}(\omega)
}{\mid 1-e^{-2
q d}R_{1p}(\omega +q_xv)R_{2p}(\omega)\mid ^2}+\left( 1\leftrightarrow 2\right) \right)
\]
\begin{equation}
\left. -\int_0^{q_xv}d\omega [n(\omega )+1/2]\left( \frac{\mathrm{Im}
R_{1p}(\omega -q_xv)\mathrm{Im}R_{2p}(\omega)}
{\mid 1-e^{-2qd}R_{1p}(\omega -q_xv)R_{2p}(\omega)\mid ^2}
+(1\leftrightarrow 2)\right) \right\}.
\label{parallel2}
\end{equation}
where $n(\omega )=[\exp (\hbar \omega /k_BT-1]^{-1}$ and ($1\leftrightarrow 2$) 
denotes the term which is obtained from the first one by permutation of indexes 
$1$ and $2$. $R_{ip}$ ($i=1,2$) is the reflection amplitude for surface $i$ for $p$ -polarized electromagnetic waves. The reflection amplitudes for a 
2D-electron system are determined by \cite{Volokitin10}
\begin{equation}
R_{ip}=\frac{\epsilon _{ip}-1}{\epsilon _{ip}+1}, 
 \label{refcoef}
\end{equation}
where $\epsilon _{ip}=
4\pi iq \sigma _i(\omega,q)/\omega \varepsilon +1$, $\sigma _i$  is the longitudinal conductivities of the layer $i$ and 
$\varepsilon$ is the dielectric constant of the surrounded dielectric. 
The longitudinal conductivity can be written in the form $\sigma _l(\omega ,q)=-\mathrm{i}\omega \chi _l(\omega
,q)/q^2,$ where $\chi _l$ is the finite lifetime generalization of the
longitudinal Lindhard response function for 2D-electron gas \cite{Mermin,Stern}.
 The friction force per unit charge in the layer is determined by 
$E=\sigma_{\|}/n_se$, where $n_s$ is the 2D-electron concentration in the
layer. For $v\ll v_F$, where $v_F$ is the  Fermi
velocity, the friction force  depends linearly on 
velocity $v$. For $d=175\ \mathrm{\AA }$ at $T=3$ K, and with
$n_s=1.5\times 10^{15}$ m$^{-2}$, the electron effective mass 
$m^{*}=0.067$ m$_e$,
$v_F=1.6\times 10^7$ cm/s,   the electron mean free path
$l=v_F\tau =1.21\times 10^5\ \mathrm{\AA }$, and $\varepsilon =10$ (which 
corresponds to the condition of the experiment \cite{Gramila1}) we get  $E=6.5\times
10^{-6}v$ V/m, where the velocity $v$ is in m/s. For a current $200$ nA in 
a two-dimensional layer with the width $w=20\mu$m the drift of electrons 
 (drift velocity $v=60$m/s) creates a frictional drag force per unit charge in 
the adjacent quantum well $E=4\cdot10^{-4}$V/m. Note that for the electron systems the frictional drag force decreases 
when the electron concentration increases. For 
a example, for 2D-quantum wells with high electron
density ($n_s=10^{19}$ m$^{-2}$, $T=273$ K, $\tau =4\times
10^{-14}$ s,  $\varepsilon =10$, $m^*=m_e$) at $d=175\ \mathrm{\AA }$ we get 
$E=1.2\times 10^{-9}v$ V/m. 

Let us replace one 2D-layer by semiinfinite chamber with liquid containing ions. The reflection amplitude for the interface between dielectric and liquid 
is given by Eq. (\ref{refcoef}), where
\begin{equation}
\epsilon=\frac{\varepsilon_0\lambda (Dq_D^2-i\omega)}{\varepsilon(Dq_D^2q-i\omega\lambda)},
\label{refcoef1}
\end{equation} 
where $D$ is the diffusion coefficient of the ions, $\varepsilon_0$ is the dielectric function of liquid without ions, $q_D=\sqrt{4\pi ne^2/\varepsilon_0 k_BT}$
 is the Debye screening wave number,
$\lambda=\sqrt{q^2+q_D^2-i\omega/D}$, $n$ is the ion concentration. In this case, for $v\ll v_F$ the friction force also depends linearly 
 on velocity $v$. 
In particular, for $n=10^{24}$ m$^{-3}$, $T=273$ K, $\varepsilon_0=80$, $D=10^{-9}$m$^2$/s for high electron density ($n_s=10^{19}$m$^{-2}$)
for 2D-electron system we get $E=1.4\times 10^{-6}v$ V/m. 
This friction is three orders of magnitude larger than the friction between the 2D-electron system 
 with high electron concentration, and of the same order 
of magnitude as the friction between 2D-electron systems 
with low electron concentration.

Let us replace the second 2D-layer  by  narrow  2D-channel
 with thickness $d_c$.  For $q_Dd_c\ll 1$ the flowing liquid in the
channel can be considered
as a 2D-liquid. The conductivity of a 2D-liquid with ions performing Brownian motion is given by
\begin{equation}
\sigma(\omega,q) = -\frac{i\omega d_c}{4\pi}\left(-1+\varepsilon_c\left(1+\frac{q_{Dc}^2}{q^2-i\omega/D_c}\right)\right)
\label{conductivity}
\end{equation}
where $\varepsilon_c$, $D_c$ and $q_{Dc}$ are the dielectric function, the diffusion coefficient and the Debye wave number for the liquid in the channel,
 respectively.
Fig. 3 shows
the dependence of the frictional drag force per unit charge on the   liquid flow velocity for identical liquid in the channel and in the 
semi-infinite chamber with the same parameters as used above for the liquid and the separation between the channel and chamber 
$d=1$nm. The frictional drag force on the ions in the channel initially 
increases with the flow velocity, reaches a maximum and then decreases at large
value of the flow velocity, in agreement with the model calculation in \cite{Das}. The position of maximum decreases when the density of
ions decreases. The
frictional drag force induced by liquid flow in the narrow channel is nine orders of magnitude larger than the friction force  induced in a 2D-electron
system. 

As a limiting case of the situation considered above, let us consider a 2D-system immersed in a flowing liquid in an infinite chamber. 
We assume that the liquid flows  along the $x$-axis, and that the plane of 2D-system coincides 
with the $xy$- plane.  According to the fluctuation-dissipation theorem 
the average value  of the correlation function for the Fourier components of the Coulombic potential for a infinite medium in the plane of 2D-system 
is determined by \cite{RMP07}
\begin{equation}
<\varphi^f (\omega,q,0)\varphi^{*f} (\omega,q,0)>=4\hbar (n(\omega) +1/2)\mathrm{Im}\Sigma(\omega,q,0),\label{one}
\end{equation}
where 
\begin{equation}
\Sigma(\omega,q,0),=-\int_{-\infty}^{\infty}\frac{dk_z}{2\pi}\frac{1}{k^2\varepsilon_0\varepsilon(k)},
\end{equation}   
where $\mathbf{k}=(\mathbf{q},k_z)$.  The dielectric function of the Debye 
plasma is determined by 
\begin{equation}
\varepsilon(k)=1+\frac{q_D^2}{k^2-i\omega/D}
\end{equation}
According to the fluctuation-dissipation theorem the average value of the correlation function for the  Fourier components of 
the fluctuating current 
density in 2D-system is determined by \cite{RMP07} 
\begin{equation}
<j_q^f(\omega,q)j_q^{f*}(\omega,q)>=\frac{\hbar q^2}{\pi \omega}(n(\omega) +1/2)
\mathrm{Re}\sigma (\omega,q).
\label{two}
\end{equation}    
The friction force per unit area of 2D-system is given by
\begin{equation}
\gamma_{\|}=\int \frac{d^2\mathbf{q}}{(2\pi)^2}\frac{q_x}{q}<
E_q(\omega,q,0)\rho^*(\omega,q)>
\label{three}
\end{equation}
where $E_q =-iq\varphi_q= E_q^f+E_q^{ind}$, $\rho=\rho^f + \rho^{ind}$.  
The fluctuating electric field $E_q^f$ is the sum of the fluctuating field 
which exists in the infinite medium without the 2D-system, and the 
electric field created by the fluctuating charge density 
$\rho_q^f=qj^f_q/\omega$ in presence of the flowing liquid. The induced 
electric field $E_q^{ind}$ is created by the charge density 
$\rho_q^{ind}=qj^{ind}_q/\omega$ induced in 2D-system by 
the electric field $E_q$. The correlation functions (\ref{one}) and (\ref{two}) are determined in the rest 
reference frames of the liquid, and 2D-system, respectively. To find the 
relation 
between electric fields in the different reference frame we use the Galilean 
transformation, what will give rise to Doppler shift of the frequencies of the 
electric fields in the different reference frames. Solving the Poisson's equation with the fluctuating charge density $\rho^f$, and 
 the charge density induced by the fluctuating electric field $E_q^f=-iq\varphi^f$ as the sources of the field we 
get $E_q$ and from Ohm low we get $\rho$. The result of these calculations is:  
\[
\sigma_{\|} =\frac \hbar {2\pi ^2}\int_{-\infty }^\infty dq_y\int_0^\infty
dq_xq_xq^2\Big\{ \int_0^\infty d\omega [n(\omega )-n(\omega
+q_xv)]
\]
\[
\times \left( \frac{\mathrm{Re}\sigma(\omega+q_xv)\mathrm{Im}\Sigma(\omega)
}{(\omega+q_xv)\mid 1-4\pi iq^2\sigma(\omega +q_xv)\Sigma(\omega)/(\omega +q_xv)\mid ^2}+\left( \omega +q_xv
\leftrightarrow \omega\right) \right)
\]
\[
 -\int_0^{q_xv}d\omega [n(\omega )+1/2] \Big( \frac{\mathrm{Re}\sigma(\omega-q_xv)\mathrm{Im}\Sigma(\omega)}
{(\omega+q_xv)\mid 1-4\pi iq^2\sigma(\omega -q_xv)\Sigma(\omega)/(\omega -q_xv)\mid ^2}
\] 
\begin{equation}
 +( \omega +q_xv
\leftrightarrow \omega)\Big) \Big\}.
\label{2DElInL}
\end{equation}
where ($ \omega +q_xv
\leftrightarrow \omega$)  denotes the term which is obtained from the first one by permutations of the 
  arguments $\omega +q_xv$ 
 and $\omega$. With the same parameters as used above for the liquid, and for the high density 2D-electron system, we get  
$E=8.1\cdot 10^{-6}v$ V/m. For a 1D-electron system we obtained a formula which is similar to Eq. (\ref{2DElInL}). Fig. 2 shows the result 
of the calculations of the friction drag force (per unit charge) for a 1D-electron system with the electron density per unit length $n_l=
3\times 10^9$m$^{-1}$, the temperature $T=300$ K, and with the same parameters for the liquid as used above. For the 1D-electron system we obtained a 
slight deviation 
from the linear dependence of the frictional drag on the  liquid flow velocity. The frictional drag for the 1D-electron system is one order 
of magnitude larger than for the 2D-electron system. 
Fig. 4 shows 
the dependence of the frictional drag force per unit charge in the the 2D-channel with liquid on the   liquid flow velocity in the infinite chamber 
assuming identical liquid in the channel and in the  
chamber. In this case the maximum in the frictional drag force is larger, and the decay at large velocities is slower in comparison with the semi-infinite 
chamber at separation $d=1$nm. 
 Qualitatively, the same results we obtained for a 1D-channel. For a channel  
with open ends  the 
frictional drag force will induce a drift motion of the ions in the liquid with 
velocity $v_d=D_ceE/k_BT$. The positive and negative ions will drift in the 
same direction. In contrast to electron systems, for the  channel with closed ends, in the case of  equal 
concentration of the positive and negative ions, the 
frictional drag will not induce a voltage because displacement of the ions under the action of the frictional drag force 
 will not violate charge neutrality of the liquid. However, the frictional drag will induce a 
pressure difference $\Delta p=nLeE$, where
$L$ is the length of the channel. For example, if $n=10^{24}$m$^{-3}$, $L=100\mu$m  
and $E=1000$V/m we get pressure difference $\Delta p=10^4$ Pa, which should be easy to measure.   

In this Letter we have shown that the van der Waals frictional drag force 
induced in low-dimensional system by liquid flow can be several orders 
of magnitude larger than the friction induced by electron current. For a narrow 2D-channel with liquid the frictional drag force is several orders 
of magnitude larger 
than for 2D-electron systems. In the contrast to 2D-electron system the frictional drag force for a narrow channel depends nonlinearly
 on the flow 
velocity. These results should have a broad application for studying of the van der Waals friction and in the design of nanosensors.

\vskip 0.5cm   

A.I.V acknowledges financial support from the Russian Foundation
for Basic Research (Grant N 06-02-16979-a), DFG and the European
Union Nanotribology project.

\vskip 0.5cm

FIGURE CAPTIONS

Fig. 1. Two bodies moving relative to each other will experience van der       
Waals friction due to Doppler shift of the electromagnetic waves emitted by 
them.

Fig. 2. The frictional drag force per unit charge in a 1D-electron system 
induced by liquid flow in infinite chamber as a function of the flow velocity.
The temperature $T=300$ K, the ion concentration in liquid $n=10^{24}$m$^{-3}$,  the diffusion coefficients of ions $D=10^{-9}$m$^2$/s, the dielectric 
constant of the liquid $\varepsilon_0=80$, the electron 
concentration per unit length $n_l=3\cdot10^9$m$^{-1}$, the electron relaxation time $\tau = 4\cdot 10^{-14}$s 

Fig. 3. The frictional drag force per unit charge for ions  in a 2D-channel
with liquid induced
by liquid flow in semi-infinite chamber as a function of the flow velocity for
identical liquids in the channel and chamber. The temperature $T=300$ K, the ion concentration in liquid $n=10^{24}$m$^{-3}$,
 the diffusion coefficients of ions $D=10^{-9}$m$^2$/s, the dielectric
constant of the liquid $\varepsilon_0=80$, the separation between the channel and semi-infinite chamber $d=1$nm.

Fig. 4. The same as Fig. 4 but for infinite chamber.

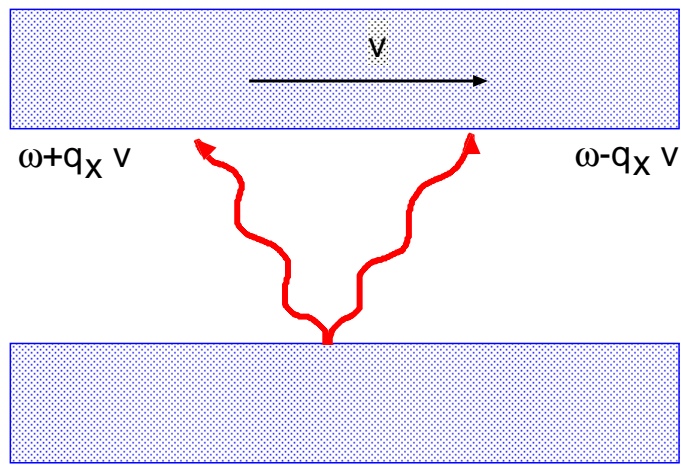
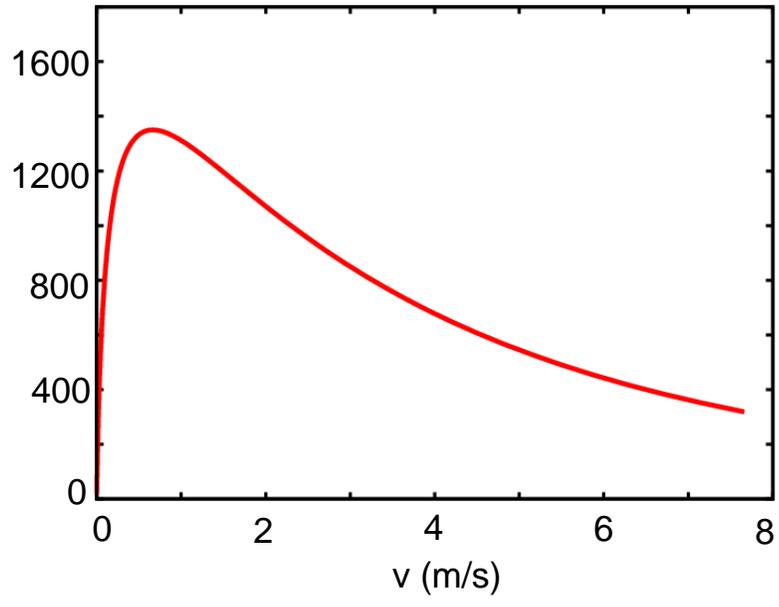
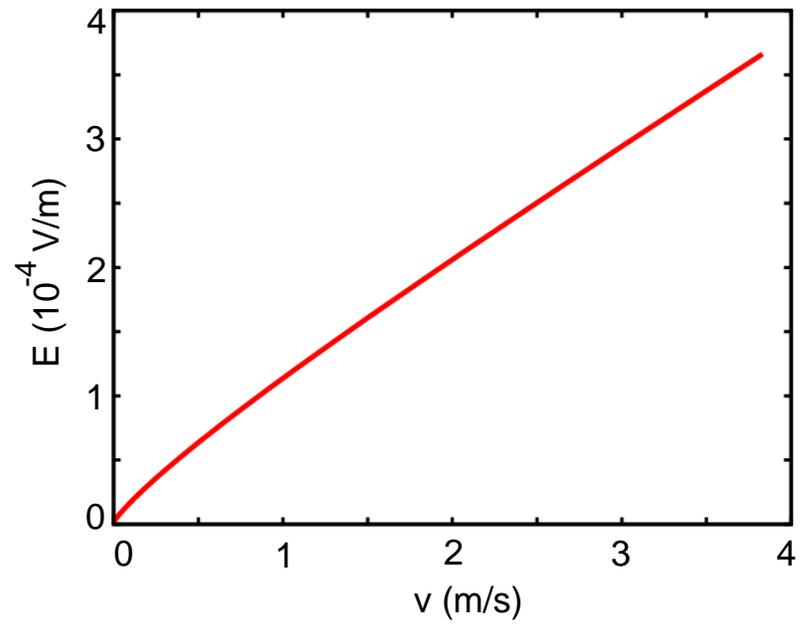
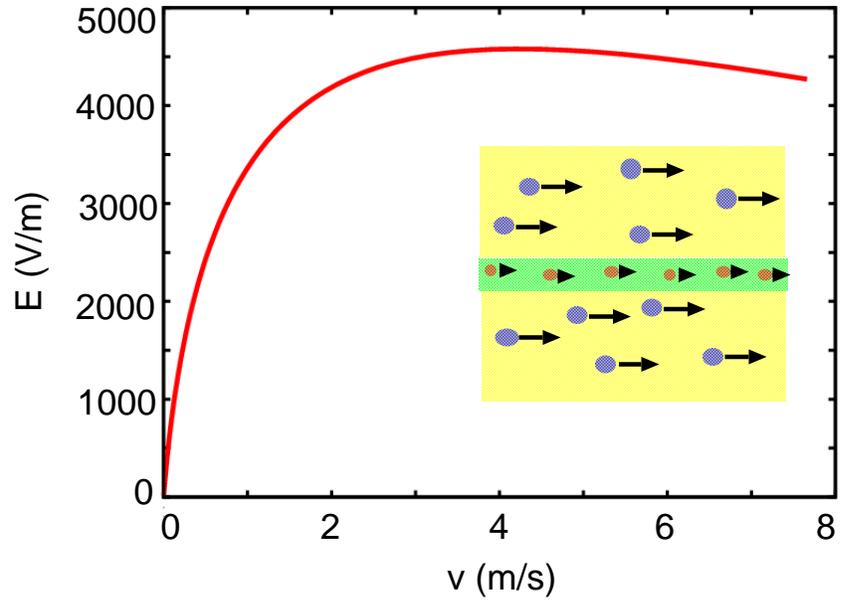
\end{document}